\documentclass[lettersize,journal]{IEEEtran}
\usepackage{amsmath,amsfonts}
\usepackage{algorithmic}
\usepackage{array}
\usepackage[caption=false,font=normalsize,labelfont=sf,textfont=sf]{subfig}
\usepackage{textcomp}
\usepackage{stfloats}
\usepackage{url}
\usepackage{verbatim}
\usepackage{graphicx}
\usepackage{balance}
\usepackage{siunitx}
\usepackage{bm}

\renewcommand{\vec}{\bm}

\begin{document}
\title{Applying Energy Absorption Interferometry to THz direct detectors using photomixers}
\author{Ian Veenendaal, Edgar Castillo-Dominguez, Stephen J.C. Yates, Bram Lap, and Willem Jellema
\thanks{Ian Veenendaal (previously), Edgar Castillo (previously), Stephen Yates, Bram Lap, and Willem Jellema are with the SRON Netherlands Institute for Space Research, Groningen, the Netherlands.}%
\thanks{Ian Veenendaal is currently with the Astronomical Instrumentation Group, Cardiff University, Cardiff, United Kingdom (email: veenendaali@cardiff.ac.uk).}%
\thanks{Edgar Castillo is currently with the Department of Astrophysics, University of Oxford, Oxford, United Kingdom.}%
\thanks{Bram Lap and Willem Jellema are also with the Kapteyn Astronomical Institute, University of Groningen, Groningen, the Netherlands.}
\thanks{© 2023 IEEE.  Personal use of this material is permitted.  Permission from IEEE must be obtained for all other uses, in any current or future media, including reprinting/republishing this material for advertising or promotional purposes, creating new collective works, for resale or redistribution to servers or lists, or reuse of any copyrighted component of this work in other works.}
}


\maketitle

\begin{abstract}
Detector requirements for far infrared astronomy generally result in devices which exhibit a few-moded response to incident radiation.
The sensitivity and spatial form of the individual modes to which such a
detector is sensitive can be determined with knowledge of the complex valued cross-spectral density of the system, which we label the detector response function (DRF).
A matrix representing the discretized cross-spectral density can be measured from the complex amplitudes of interference fringes generated by two identical sources as they are independently scanned through the field of view. We provide experimental verification of this technique using monochromatic THz beams generated by photomixers in which the relative phase is varied with fiber stretchers. We use this system to characterize the modal response of a single pixel from an array of microwave kinetic inductance detectors (MKIDs). 
\end{abstract}

\begin{IEEEkeywords}
  Energy Absorption Interferometry, photomixer, partial coherence, verification techniques
\end{IEEEkeywords}

\section{Introduction}
\IEEEPARstart{T}{he} next generation of far-infrared (FIR) space-based observatories will incorporate cold primary optics and ultra-sensitive superconducting detectors with NEP on the order of \num{e-20} \unit{W/\sqrt{Hz}} \cite{baselmansKilopixelImagingSystem2017, baselmansUltrasensitiveTHzMicrowave2022}.

These observatories will seek to achieve unprecedented spectral-spatial resolution in order
to, for example, better understand the minutia of star and planet formation within our own
galaxy \cite{farrahReviewFarinfraredInstrumentation2019}.

Resolution, throughput, and bandwidth requirements for these wavelengths put strict limitations
on the size and density of detectors \cite{wuObservingExtendedSources2013, shuPrototypeHighAngular2018}. 
It is generally that case that these requirements drive the optimum size of detectors
to be similar to the wavelengths measured, such that they exhibit a few-moded
response to incoming radiation.
In such a regime, the detector beam pattern cannot be fully represented by a fully 
coherent response, as is the case for single-moded detectors. In addition, the system
also cannot be fully described using geometric optics and radiative transfer models, 
such as typically used in the infrared and visible domains.

Instead, the THz detector field is partially coherent, being
described as an incoherent linear superposition of a set of individually
coherent fields, which are understood to be the natural modes of the
system. Validating the performance of few-moded power absorbing detectors
requires a method to measure the number and relative sensitivities of the
natural modes, as well as their spatial forms.

In addition, the number of modes present and relative sensitives of a system to
a defined mode set is frequency dependent. Therefore, it is beneficial to be
able to measure these properties over an arbitrary set of frequencies.
Continuously tunable THz sources, such as photomixers, have been shown to be
particularly valuable as test sources for experimental measurement and
verification of the frequency-dependent modal response of detectors \cite{Roggenbuck2012}.

In this work, we applied Energy Absorption Interferometry (EAI), which is a general
technique by which the coherent properties of any power absorbing structure can
be measured, to an end-to-end THz optical system including detectors \cite{withingtonProbingDynamicalBehavior2012}.
In doing so, we demonstrated we were able to recover a complex valued correlation
matrix which described sensitivity of the system to individually coherent natural modes.

We introduced two independent source
probes in the field of view of an optical system, which while being spatially
decoupled, exhibited identical beam profiles, were frequency matched, and the
relative phase between the sources could be adjusted. The sources were
independently scanned throughout the field of view, and at each pair of
positions, the relative phase was swept and an interference fringe was measured.

Extracting the phase and amplitude of the fringe allowed us to construct a
directly measured representation of the spatial correlation matrix of the
complete optical system.  Such a matrix could be propagated through the optical system, and
decomposed into a weighted set of orthogonal functions which represented the individually coherent
optical modes of the system.

\section{Theory}
\noindent
Uncovering the natural modes of a system can be accomplished by
measuring the second-order correlations within the system, using a
technique known as Energy Absorption Interferometry (EAI) \cite{Withington2017}.

\subsection{Optical Modes}

A partially coherent, monochromatic beam can be represented by an ensemble of linear superpositions of a set of spatially coherent propagating fields \cite{wolfNewTheoryPartial1986},

\begin{equation}
    V(\vec{r}, \nu) = \sum_{n}^{} a_n(\nu) \psi_{n}(\vec{r}, \nu).
    \label{eqn:beam}
\end{equation}

where $\vec{r}$ is the position vector of a point in the beam, $\nu$ is the optical frequency. $a_n(\nu)$ are random complex values which represent fluctuations in amplitude and phase over the ensemble, and $\psi_n(\vec{r}, \nu)$ are orthogonal functions which represent self-coherent but mutually incoherent spatial modes. In this work, we consider only monochromatic sources, and therefore the explicit frequency dependence is dropped in the following expressions.

Spatial correlations may be represented by the cross-correlation function of the beam at two points, $\vec{r}_1$ and $\vec{r}_2$ \cite{wolfNewTheoryPartial1982},

\begin{equation}
    \langle V^{\ast}(\vec{r}_1) V(\vec{r}_2)\rangle = \sum_n \sum_m \langle a_n^{\ast} a_n \rangle \psi^{\ast}(\vec{r}_1) \psi(\vec{r}_2),
    \label{eqn:cross-correlation}
\end{equation}

where the asterix defines the complex conjugate and the angle brackets denotes the average over an ensemble of beams. We assume here that our sources and optical systems are stationary and with narrow random fluctuations, such that the ensemble average can be represented by a time average.

In our measurements, we cannot measure the instantaneous and complex values of $a_n$, but we can identify their statistical properties. We define the second-order moments as,

\begin{equation}
    \langle a_n^{\ast} a_n \rangle = \lambda_n \delta_{nm},
\end{equation}
where $\lambda_n$ is a constant and $\delta_{mm}$ is the Kronecker delta.

Equations \ref{eqn:beam} and \ref{eqn:cross-correlation} can be used to describe the partially coherent output beam of a source or, using the principle of reciprocity, the partially coherent reception pattern of a detector. In addition, an optical system placed in front of a detector will modify the coherent properties of the detector's reception pattern in a repeatable way. Therefore, we can expand our definition of a detector to include all preceding optics between the detector and a source, and consider the partial coherence of the reception pattern of the entire optical system.
In this work, we will distinguish between the optical modes of a source and the optical modes of a receiving system (including detector and preceding optics) by labelling them optical modes and detector modes respectively.

\subsection{Cross-Spectral Density}
\label{subsec:csd}

The expression in Equation \ref{eqn:cross-correlation} is an important property of a partially coherent beam, which is know as the cross-spectral density $W(\vec{r}_1, \vec{r}_2)$, which is in our case evaluated only at a single frequency, $\nu$. The cross-spectral density is itself the frequency Fourier transform of the mutual coherence function at two spatial points and two times, $\Gamma(\vec{r}_1, \vec{r}_2, t_1, t_2)$ \cite{mandelOpticalCoherenceQuantum}.
Wolf shows that the previously defined $\psi_n$ and $\lambda_n$ are eigenfunctions and eigenvalues of the following equation \cite{wolfNewTheoryPartial1982},

\begin{equation}
    \int_{D} W(\vec{r}_1, \vec{r}_2) \cdot \psi_n (\vec{r}_1) d^3\vec{r}_1 = \lambda_n \psi_n(\vec{r}_2)
    \label{eqn:fredholm_integral}
\end{equation}

where the integral over $D$ represents integrating over all positions of $\vec{r}_1$ in the domain $D$ in which we want to define the modes $\psi_n$. An important takeaway here is that if the complex-valued cross-spectral density of a system can be extracted through measurements, we can find a set of eigenvalues and eigenfunctions which can be used to describe the beam we are trying to measure. In most cases, we are interested in the intensity properties of the beam, in which case only $\lambda_n$ is needed, and not $a_n$. In addition, the number of unique eigenvalue and eigenfunctions determined is used to describe the number of modes of the beam.
If a single eigenvalue and eigenfunction can be found, then the beam can be described as single-moded. If more than one are found, the beam is multi-moded.

\subsection{Detector Response Function}

Moving to the domain of experimental techniques, in order to measure the beam profile of an optical system (including a receiver), we must illuminate it with a source and measure its response. To simplify the terminology here, we will refer to the entire optical system and receiver as the detector. The source will itself have a beam, which may be single-moded or multi-moded, such that what is measured is the convolution of the optical modes of the source and the modes of the detector. Both source and detector will have their own cross-spectral density functions, which we label $E$ and $D$.

Withington et al. show that the power coupling of the system is a projection of the cross-spectral density of the detector beam, $\overline{\overline{D}}(\vec{r}_1,\ \vec{r}_2)$, as well as the cross-spectral density of the source field, $\overline{\overline{E}}(\vec{r}_1,\ \vec{r}_2)$ \cite{Withington2017}. 

\begin{equation}
    \langle P \rangle = \iint_{S^{2}}{
    \overline{\overline{D}}\left(\vec{r}_1,\vec{r}_2 \right)
    \cdot \cdot
    \overline{\overline{E}}\left( \vec{r}_1,\vec{r}_2 \right)
    d^2 \vec{r}_1 d^{2}\vec{r}_2 }.
\end{equation}

The $\overline{\overline{}}$ emphasises these quantities as second-order tensors defined continuously over $\vec{r}_1$ and $\vec{r}_2$, and $S$ is a plane over which power-absorbing nature of the detector is determined. In the following sections, we will use a more familiar matrix representation.
In addition, keeping with the terminology in \cite{Withington2017}, the cross-spectral density of the detector beam is defined as the two-point detector response function (DRF).
We also select our choice of sources carefully in order to simplify the expression for $\overline{\overline{E}}$ so that the DRF can be easily extracted from the measurements.

\subsection{Matrix Representation}
Since experimental application relies on individually sampling discrete position of the source probes, we used these spatial coordinates at which the source probe is positioned to define the basis over which
$\overline{\overline{D}}\left(\vec{r}_1,\vec{r}_2 \right)$
and
$\overline{\overline{E}}\left( \vec{r}_1,\vec{r}_2 \right)$
were measured.
In this work, we consider only a one-dimensional slice of the detector beam which we aim to characterize, to allow for a simple matrix representation. However, expansion for a measurement of a cross-sectional plane of the detector beam is also possible and will be the goal of future work.

We also aim to define the positions of the source probes at the coordinates determined by the desired sampling size and resolution of the detector beam at a focal plane,
such that both
$\overline{\overline{D}}\left(\vec{r}_1,\vec{r}_2 \right)$ and $\overline{\overline{E}}\left( \vec{r}_1,\vec{r}_2 \right)$
can be measured in the same basis,
and no change of basis operations are required.
The definition of
$\overline{\overline{D}}\left(\vec{r}_1,\vec{r}_2 \right)$ and $\overline{\overline{E}}\left( \vec{r}_1,\vec{r}_2 \right)$
in arbitrary locations along the beam propagation axis is possible and is detailed in Moinard et al.~\cite{Moinard2017a}.
This sampling coordinates can be expressed as a set of $M$ $x$-coordinates ${x_i}$ where $i\in{1, 2, ..., M}$. These can in turn be expressed as a set of basis vectors ${\vec{x}_i}$ where each $\vec{x}_i$ contains a single non-zero element.

Taking the possible sampling coordinates of the detector beam to be $x_i$ and the position of a source probe to be and index $m$ ranging from $1$ to $M$, we can express a discretized optical field of a source probe as a vector in the basis of $\vec{x}_i$ as,

\begin{equation}
    \vec{e}_m = \sum_i a_{i, m} \vec{x}_i,
    \label{eqn:discrete_single_source}
\end{equation}
where $a_{i, m}$ is a complex value representing the amplitude and phase of the source field at detector sampling coordinate $x_i$ by a source probe located at position $m$.

The EAI technique relies on using two identical source probes to map correlations of the detector beam, therefore we will consider the field produced by two such sources as described by Equation \ref{eqn:discrete_single_source}.
If two such identical sources are positioned at indices $m$ and $m^{\prime}$, the resulting optical field produced by both sources can be expressed as,
\begin{equation}
    \vec{e}_{m, m^{\prime}} = \vec{e}_m + \vec{e}_{m^{\prime}} \exp{\left(i \phi\right)}
    \label{eqn:discrete_dual_source}
\end{equation}
where $\phi$ is an intrinsic phase difference between the two sources. The two source field cross-spectral density matrix is then,
\begin{equation}
    E = \langle \vec{e}_{m, m^{\prime}}^{\dagger} \vec{e}_{m, m^{\prime}} \rangle
\end{equation}

We drastically simplify the discretized representation again by considering the case where the source probes are point-like sources situated at the detector sampling coordinates $x_i$. In this case, there is only a single non-zero $a_{i, m}$ in Equation \ref{eqn:discrete_single_source} for each source position $m$. In this way, the source fields $\vec{e}_m$ are linearly independent vectors.

For a particular pair of source probe locations $m$ and $m^{\prime}$, the power measured at the detector can be express as \cite{Moinard2017a},

\begin{equation}
    \left\langle P(\phi) \right\rangle = D_{m, m} + D_{m^{\prime}, m^{\prime}} + 2\ \left| D_{m, m^{\prime}} \right|\cos\left( \phi + \theta_{m,m^{\prime}} \right).
    \label{eqn:dnnp}
\end{equation}

Here, $D_{x, y}$ are individual elements of the DRF, $D$, in the basis of the source positions, 
$D_{n,n}$ and $D_{n^{\prime},\ n^{\prime}}$ are the detector response
to the individual source beams at positions $n$ and $n^{\prime}$, and
$\left| D_{n,\ n^{\prime}} \right|$ and $\theta_{n,n^{\prime}}$ are the
amplitude and phase of the complex degree of coherence of the system
response to the two interfering sources at $n$ and $n^{\prime}$.

The amplitudes of the field $a_n$ and $a_{n^{\prime}}$ are absorbed into the definition of $\left| D_{n,\ n^{\prime}} \right|$ in  Equation(\ref{eqn:dnnp}). Although not strictly correct, our interest was only in the relative spatial forms and amplitudes of the detector modes, and therefore these constant field amplitudes were removed for simplicity.

By rotating the relative phase difference $\phi$ between the sources the system response will trace out an interference fringe. The amplitude and phase of the fringe pattern can be extracted if the incident phase difference $\phi$ is known.
Furthermore, by independently scanning the two sources over all possible pairs of coordinates $x_n$,
all elements of the DRF are measured.

With the entire DRF defined, we can return to the claim made in Section \ref{subsec:csd} in which, knowing the cross-spectral density of the detector beam, we can recover the detector modes and relative sensitivities.

\section{Setup}
\subsection{Sources}
In order to accurately recover the cross-spectral density matrix which
represents the state of coherence of the system under test, the source probes used should
be: identical, self-coherent, phase-locked, and with a well-defined
polarization. In addition, one would typically like to measure the state of coherence of the system under test at various frequencies, and therefore arbitrary frequency tunability is another desirable property of the sources.

\subsubsection{Source Selection}
There are a number of viable THz sources that were able to meet the above requirements that were considered for use in the experimental setup.
The two main continuous wave monochromatic sources under consideration were frequency multiplier chains (such as used in VNA extension modules) with waveguide probes \cite{croweVNAFrequencyExtenders2011}, or lens-coupled photoconductive antennas (PCAs), commonly known as 'photomixers' \cite{Preu2011}. Although the VNA would provide a significantly higher output signal levels, and therefore signal-to-noise, photomixers provide a higher upper frequency limit with sufficient signal level for ultra-sensitive detectors intended for astronomical instrumentation. In addition, the physical size of the VNA extension heads in comparison to the compact photomixers make the source probe scanning more cumbersome. 
Once the source probes were selected, the choice must be made to utilize two phase-locked source probes, or a single probe with a 50:50 beamsplitter and a delay line to produce two similar probe fields with a variable phase offset. Although a single probe reduces setup cost and avoids the need for additional phase-locking, this technique would require additional collimating optics, or correction of the mismatched beams due to the optical path difference from the delay line.

\subsubsection{Photomixers}
In answer to the tradeoffs presented above, we selected two identical GaAs photomixers which had been previously procured from Bakman Technologies
\cite{PB1319Photomixers}. The photomixers were fiber-fed by a pair of
frequency-tunable near infrared (NIR) diode lasers, coupled together with a 50:50 single-mode fiber
splitter. The two lasers fed into the two input ports of the splitter, and the two output ports fed into the two photomixers.
The output frequency of the photomixers was the difference
frequency of the two lasers and the optical path length to each photomixer
was designed to be identical, such that the output of the two photomixers was
time-coherent. To measure the complex amplitude of the resulting interference fringe, we required that the relative phase between the sources was adjustable. To this
end, a fiber stretcher was placed in the fiber path to each source at each splitter output port, such that
the fiber feeds could be independently stretched to vary the optical path length
to each source. 

\subsubsection{Laser Diodes}
The sources were driven by the same pair of lasers, coupled to the fiber splitter with equal path lengths, so that their outputs were monochromatic and phase-locked.
A pair of Distributed Bragg Reflector (DBR) lasers (Photodigm X-Mode \cite{incPhotodigmXModeDBR}) with center wavelengths of \qty{767}{\nm} and \qty{770}{\nm} were used to drive the photomixers.
The lasers were temperature controlled with a pair of laser and thermoelectric cooler (TEC) controllers (Arroyo 6300 Series ComboSource \cite{6300SeriesComboSource}), with a temperature precision of \qty{< 4}{\milli\K} and a current precision of \qty{< 1}{\uA}, allowing for a wavelength stability of \qty{< 0.3}{\pm}.
The selected lasers each exhibited a mode-hop free tuning range of \qty{2}{\nm},
and tuning rates of \qty{56}{\nm\per\kelvin} and \qty{1.5}{\nm\per\mA}.
With temperature limits of \qtyrange{10}{40}{\degreeCelsius},
the achieved frequency (wavelength) range was \qtyrange{0.835}{2.72}{THz} (\qtyrange{360}{110}{\um}).

\subsubsection{DC Bias}
A custom bias circuit was designed and fabricated to provide a maximum \qty{24}{\volt} DC bias to each photomixer. The bias to each photomixer was independently voltage adjustable via a potentiometer controlling dial on the front of the unit, and both the output voltage and current were displayed. Since the output THz beam power depends on the bias voltage \cite{Preu2011}, this allowed for a user to equalize the output power of both sources with dissimilar fiber coupling and THz generation efficiencies. This would be done to maximize the fringe amplitude when measuring the DRF.

The bias incorporated a safety latch which cut power to the devices if the bias current exceeded \qty{0.4}{\uA}. In this way, the maximum rated current of \qty{0.5}{\uA} was never reached. 

\subsubsection{Polarization}
The photomixers included a square spiral antenna, which produced a right-handed circularly polarized output THz beam. Polarizers could be placed between the source and the coupling optics to generate a desired state of linear polarization.

\subsubsection{Spatial Filtering}
An important aspect of recovering the optical modes of the system under test is
accurate knowledge of the source beam. In addition, the far-field divergence
angle of the source beam should be equal to or greater than that of the system
under test to ensure appreciable coupling over the scan range.

\begin{figure}
    \centering
    \includegraphics[width=0.8\linewidth]{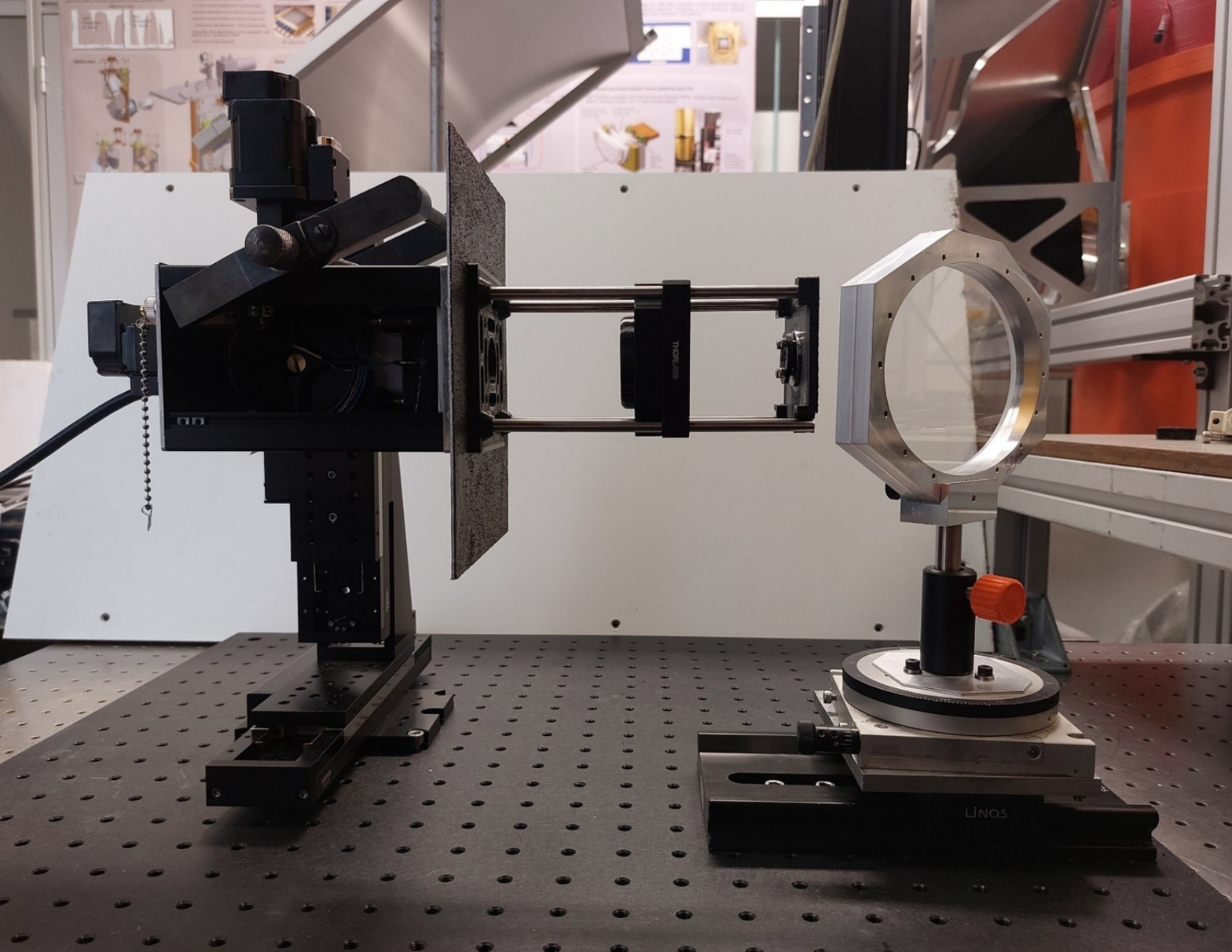}
    \caption{Image of the enclosure in which one photomixer was mounted, as well as the beam-forming optics (not including polarizer). The beamsplitter is also shown in the right hand side of the image. Also pictured are the absorber panels positioned between the photomixer enclosure and the beam-forming optics.}
    \label{fig:sources}
\end{figure}

The photomixers included an integrated hyper-hemispherical silicon lens, producing a highly collimated output THz beam.
However, due the the square spiral form of the antenna, the output beam also exhibited a frequency dependent ellipticity.
In order to be suitable for EAI, we coupled each source to an HDPE lens and placed an aperture at the produced focus to
achieve a point-like source near the focus of the system under test (Figure \ref{fig:sources}).
We used \qty{30}{mm} focal length lenses, with a \qty{3}{mm} radius near field aperture on the photomixer, and a \qty{1}{mm} radius aperture at the output of the lens, also in the near field. This spatially filtered the beam from an elliptical beam with a \qty{4}{mm} by \qty{2}{mm} waist radius, to a \qty{0.7}{mm} symmetric beam waist \cite{Goldsmith1998}. As this was smaller than the device under test (\qty{1.8}{mm} beam waist as measured) by more than a factor of two, the spatial forms of the beams measurements were dominated by the detector beam.

The aperture sizes presented above were selected in a trade-off between power coupling to the detector and broadening of the measured detector beam due to effects of the finite source probe size. Since the priority of this setup was to demonstrate the principles of the EAI measurement technique, a factor of two difference between the source beam and detector beam was considered sufficient. Further implementations of this technique may use smaller relative aperture sizes and sources with a higher output power to provide more accurate measurement of the spatial forms of the detector beam modes. If necessary, full characterisation and removal of source beam can be done \cite{Moinard2017a}, but is beyond the scope of this work.

The photomixers were placed in a custom-made enclosure in order to
\begin{itemize}
    \item allow convenient and repeatable mounting to a scanner system
    \item protect the delicate fiber attached to the photomixer itself, allowing for a ruggedized fiber to be used for connections from the scanner to the lasers.
    \item allow mounting of apertures in from the the photomixer to optically ``clean'' the output beam
    \item allow for absorber panels to be mounted around the photomixer to mitigate standing waves and give a constant background for the device under test.
    \item include a \qty{30}{mm} rail system onto which lenses, polarizers, apertures, and other optical elements could be mounted.
\end{itemize}

\subsubsection{Beamsplitter}
A beamsplitter comprised of \qty{15}{\um} thick Mylar was used to place an image of both sources in the same plane in front of the device under test \cite{naylorMylarBeamsplitterEfficiency1978}. The polarization-dependence of the transmission:reflection ratio was determined, and the beamsplitter was oriented such that a 50:50 transmission:reflection ratio was achieved for the desired state of linear polarization being measured. Note that the amplitude of DRF matrix element is proportional to the product of the source field amplitudes. Therefore, it is possible to extract the DRF with mismatched source amplitudes, albeit with a reduced fringe amplitude.

\subsection{Scanning}
The elements of the DRF were extracted by independently positioning the photomixers and accompanying beamforming optics with motorized XYZ stages. The relative phase between the sources was varied with fiber stretchers.

\subsubsection{Fiber Stretchers}
Phase rotation was achieved with \qty{15}{m} coils of single-mode, polarization
maintaining optical fiber wrapped around piezo-actuated fiber stretchers
\cite{Roggenbuck2012}.
Two stretchers were used, one for each photomixer source, which modified the
differential length between the optical paths, and kept the total optical path
within each fiber from laser to photomixer as close as possible, to minimize
frequency errors caused by thermal drifts.
The fiber stretchers used (Optiphase PZ3-PM2) gave a stretch factor of \qty{1.3}{\um\per\volt} for an optical path displacement of \qty{1.9}{\um\per\volt}.
The fiber used was Corning PM 850, designed for single-mode operation from \qtyrange{780}{900}{\nm}.
A high voltage piezo driver (Texas Instruments DRV2700EVM-HV500) was fed with a
\qty{2}{Vpp}, \qty{80}{Hz} sawtooth wave from an Agilent 33522A Arbitrary Waveform Generator, and output an AC signal with \qty{300}{Vpp}. A sawtooth wave was chosen since the phase advancement $\phi$ exhibits a linear dependence on optical path length, $\Delta L$,

\begin{equation}
    \phi = \frac{2 \pi \Delta L}{\lambda}
\end{equation}

In this way, the measured power during a linear
region of the sawtooth wave exhibited the typical cosine fringe in the time
domain (Equation \ref{eqn:dnnp}), from which the amplitude and phase could be easily extracted.

\subsubsection{Motorized Stages}
Each source is independently positioned and scanned in front of the
system using 3 stepper-motorized stages with \qty{100}{\mm} of travel (Standa 8MT175-100) in the X and Y directions, and \qty{30}{\mm} of travel (Standa 8MT173-30) in the Z direction. The stages provided a \qty{2.5}{\um} (\qty{1.25}{\um}) positioning resolution in the XY(Z) directions in full step mode. A higher positioning resolution was achievable with 1/8th microstep capabilities, but was not utilized here.

Since phase rotation was achieved by stretching of the optical fibers, the
Z-stages were not strictly needed, but provided a means of fine-tuning the positions of the sources in the same cross-sectional plane of the beam.
Additionally, first order standing wave correction was implemented by measuring the DUT twice, with the z-position varied by a quarter wavelength.
The two measurements were phase corrected and averaged, removing the dominant standing wave component caused by reflections off the source.

\begin{figure*}
    \centering
    \includegraphics[width=0.75\linewidth]{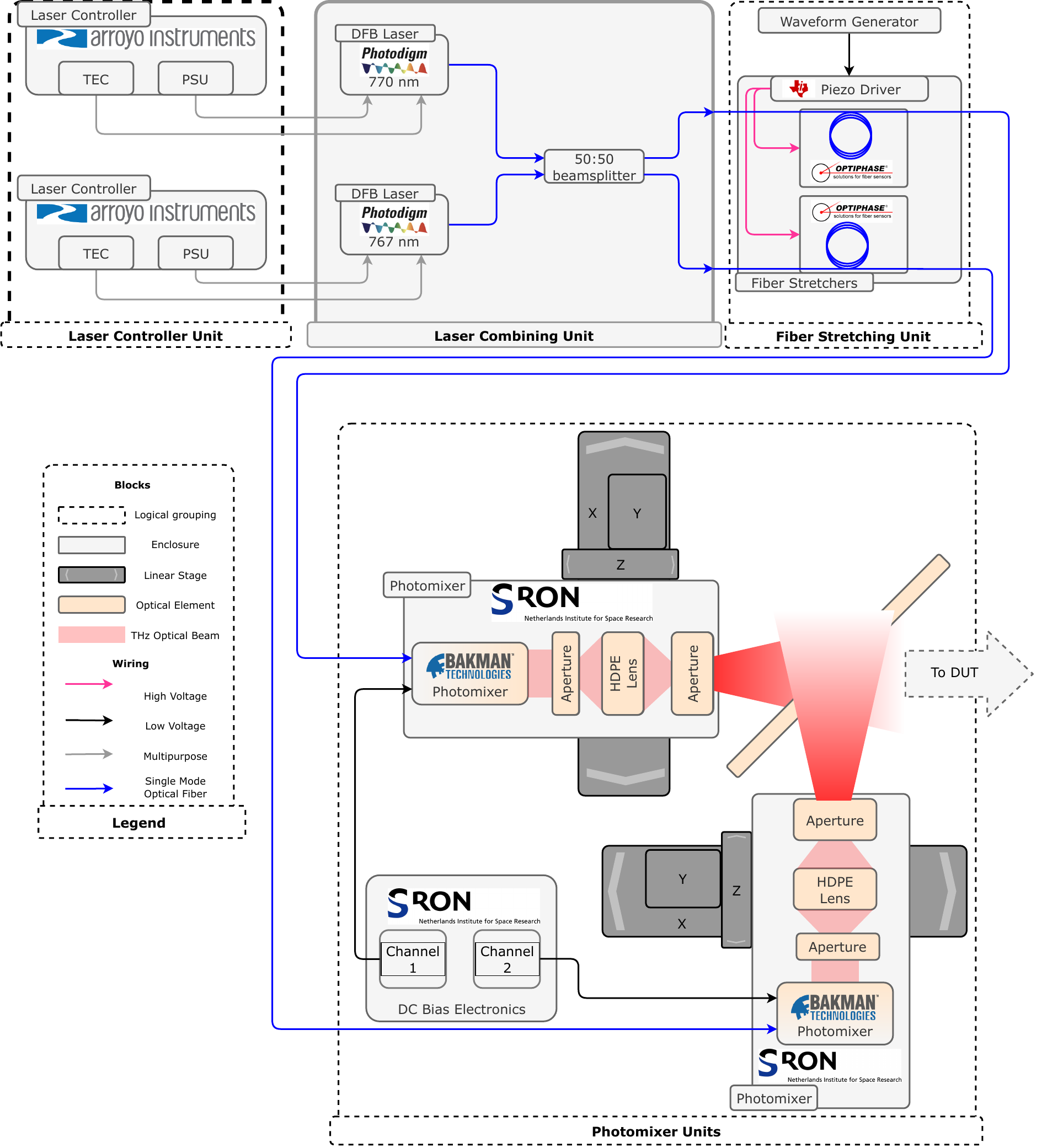}
    \caption{Schematic of the experimental setup consisting of photomixer sources, fiber stretchers, motorized linear stages, and beam-forming refractive optics.}
    \label{fig:schematic}
\end{figure*}

Including the sources, bias circuitry, and scanning electronics, the final schematic can be seen in Figure \ref{fig:schematic}

\subsection{Test System}
The technique is validated by testing a single pixel from large array (5500
pixels) of \qty{1}{mm} spaced lens-antenna coupled MKIDs optimised for 850GHz.
The array is similar to Baselmans et al.
\cite{baselmansKilopixelImagingSystem2017}, consisting of a hybrid NbTiN-Al MKID
coupled to a twin-slot antenna, but lacking the stray light absorber. Efficient
radiation coupling to free-space is achieved using lens arrays of elliptical Si
lenses on the back side of the pixel array, equipped with $\lambda/4$
anti-reflection coatings.

The system was tested using a single pixel triggered readout, conceptually similar to Day et al. \cite{dayBroadbandSuperconductingDetector2003}, and similar performance has been achieved by careful synchronization of the phase modulation with a multiplexed readout \cite{reyes2023}. The full system was tested in the lab-based wide field camera \cite{ferrariMKIDLargeFormat2018}, near the warm reimaged focal plane.

This system was chosen because of both the availability of sensitive detectors to use with the low power photomixer sources, and the well-understood nature of the optical
system, having been simulated and produced verifiable results matching simulation \cite{Davis2019, yurduseven2022}.

As verified by simulations, the MKID was primarily sensitive to a single optical mode. However, the antenna were also primarily sensitive to a single linear polarization, and we expected to encounter a maximum cross-polarization signal at the \qty{-35}{dB} level as reported in \cite{Davis2019} for a similar device with the same cryostat optics.

In addition, the lack of straylight absorber in this particular MKID array
indicated that radiation which was not absorbed by a detector could be scattered
in such a way as to create a surface wave inside the detector chip. This wave
could be absorbed by other pixels in the array. This absorbed straylight was
expected to be coherent with the main beam, resulting in an enhanced signal level away from the center of the beam in the dominant mode when comparing measurements to instrument simulations.

\subsection{Modulation and Readout}
The fiber stretchers were modulated with an \qty{80}{Hz} asymmetric sawtooth
wave. This waveform generated a series of linear voltage ramps, comprising 85\% of the waveform period, and a short time to reset to the original position in the remaining 15\% of the period. This waveform was chosen to provide a
linear variation in phase during each source position pair, which
simplified extraction of the amplitude and phase of the corresponding
fringe in the frequency domain, as well as a conservative settling time to allow the fibers to return to their original positions.

\begin{figure}[!t]
    \centering
    \includegraphics[width=0.9\linewidth]{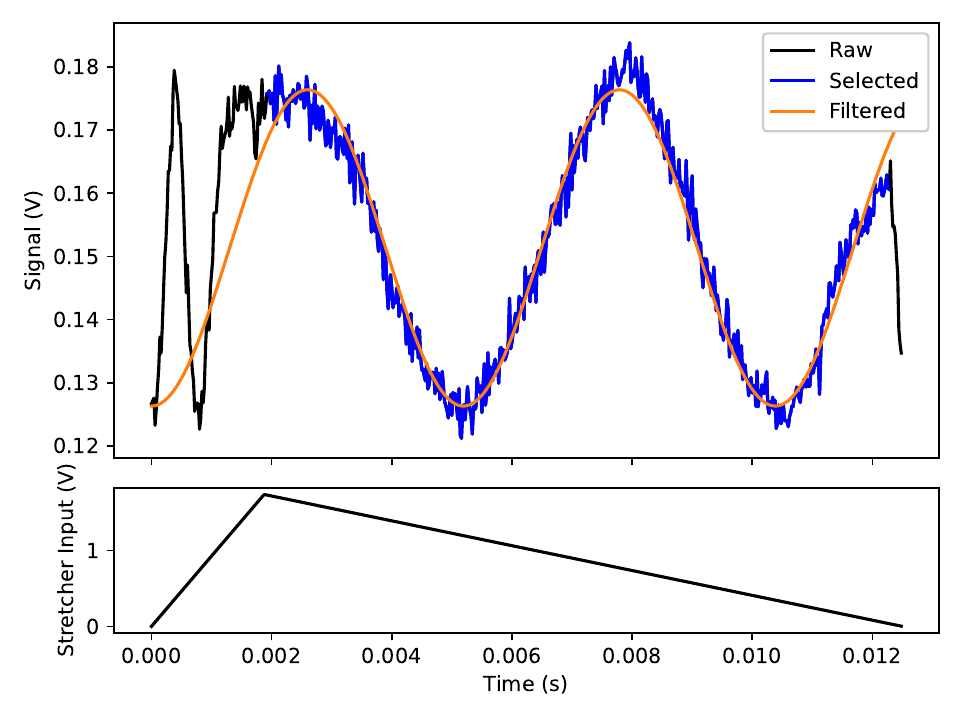}
    \caption{A representative measurement buffer from which the complex-valued fringe amplitude could be extracted. The top figure shows a timestream of the detector readout, with the data windowed for fitting and a best fit sinusoid to the windowed data. The bottom figure shows the voltage applied to the fiber stretcher over the same time period, tracing out an asymmetric sawtooth wave. The extracted phase and amplitude of the sinusoidal fit constituted one element of the DRF matrix.}
    \label{fig:fft_buffer}
\end{figure}

The KID readout was converted to an effective KID frequency shift using the calibration scheme in Bisigello et al. \cite{bisigelloCalibrationSchemeLarge2016}.
The readout was configured for a \qty{50}{kHz} sample rate, storing data in a
buffer consisting of 625 points. The acquisition was triggered by square wave
pulses synchronized with the fiber stretcher signal. In doing so, one complete
buffer coincided with a single period of the fiber stretcher. 32 consecutive
buffers were averaged. A typical averaged buffer is shown in the top frame of Figure
\ref{fig:fft_buffer}. This resulted in an effective measurement speed of \qty{0.4}{s}
per interference fringe. 

To extract the amplitude and phase of the measured fringe in the buffer, the portion of the buffer corresponding to the 15\% reset time was first removed. Then an FFT of the remaining signal was taken, and the complex value of the bin corresponding to the expected fringe frequency was recorded.

For our test case, we operated the photomixers with a THz frequency of \qty{870}{\GHz}.
A change in optical path of \qty{3.8}{\um\per\volt} and a stretcher voltage range of \qtyrange{0}{150}{\volt} in 85\% of a period of an \qty{80}{\Hz} waveform resulted in an expected fringe frequency of \qty{155}{\Hz}.

Figure \ref{fig:fft_buffer} shows an example of the output
fringe from the detector readout, the input signal to the stretcher, and
the processing s.eps.pdf applied to the detector readout.

\section{Results}
\subsection{Complex Beam Map (2D)}

\begin{figure}[!t]
    \centering
    \includegraphics[width=0.9\linewidth]{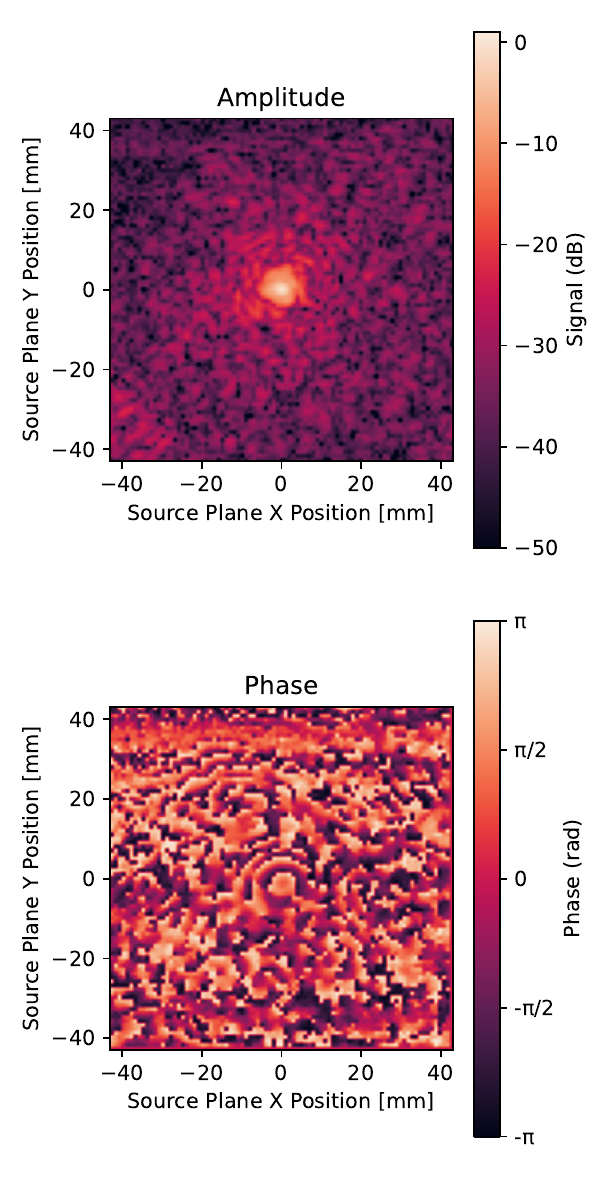}
    \caption{A two-dimensional complex-valued beam map of the detector retrieved by holding one source stationary and rastering the other over a measurement plane.}
    \label{fig:raster_beam}
\end{figure}

The source probes and their scanning configuration was versatile enough to be used to demonstrate multiple methods of beam mapping. In addition to the intended method of scanning all possible two-source position pair combinations to recover two-point correlations, one source probe could be held stationary at the center of the optical axis while the second probe scanned. An interference fringe could be measured at each position of the second probe, and by mapping the complex amplitude of the interference fringe to a two-dimensional array of source probe points which constitute a measurement plane, the complex-valued beam pattern of the convolved source probe beams and detector beam could be measured. Approximating the source probes as point sources in the measurement plane, we could retrieve the complex-valued detector beam in the measurement plane. This approach is the homodyne analogue to the heterodyne amplitude and phase measurement technique for characterizing single-mode system described in Davis et al. \cite{Davis2019}, which shows that EAI is a generalized form of the classic phase and amplitude beam pattern measurements described therein. Figure \ref{fig:raster_beam} shows a measured complex-valued beam map of the detector sampled at the source scanning plane.

Since phase information is retrieved using this technique, the measured beam pattern could be propagated forwards or backwards along the optical axis using, for example, a technique based on the angular plane wave spectrum (APWS) representation of the beam \cite{Davis2018, martinLongwaveOptics1993}.

\subsection{DRF Measurement (1D)}

\begin{figure}[!t]
    \centering
    \includegraphics[width=0.9\linewidth]{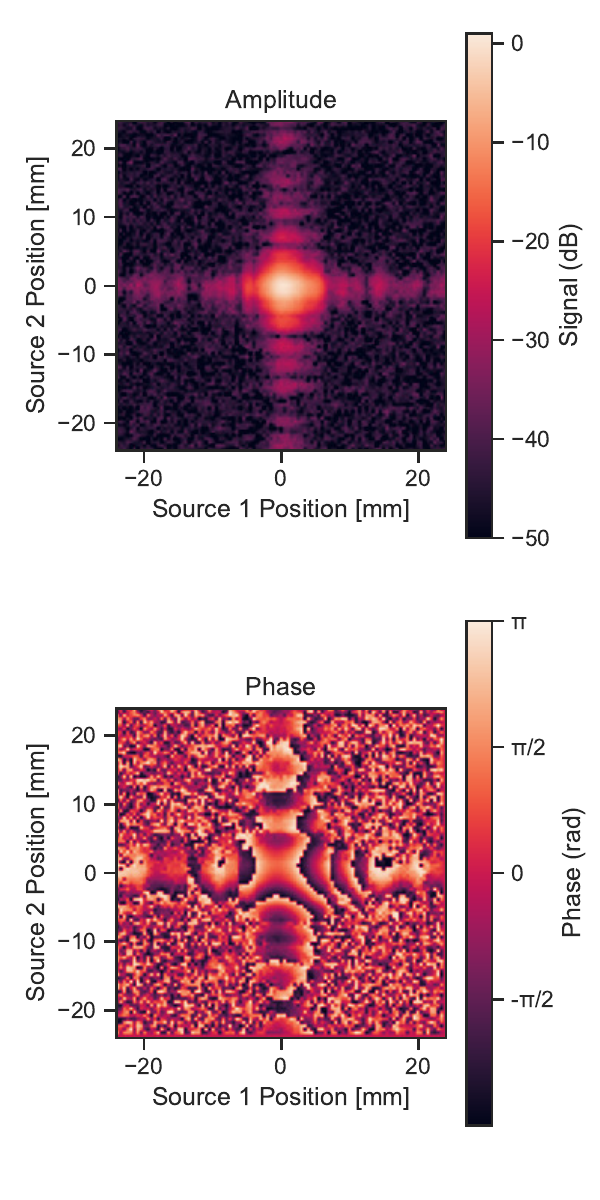}
    \caption{Measured amplitude and phase of complex valued detector response function (DRF).}
    \label{fig:drf}
\end{figure}

The detector response function is produced by recording the amplitude
and phase of the fringe pattern for each source position pair. The
amplitude and phase is shown in matrix form in Figure \ref{fig:drf}.

We limited the source position points to include only a single cut of the focal
plane, but at a diagonal between the E- and H- planes such that the cut
intersected non-zero regions of both the simulated co- and cross- polarization
responses of the system.

\subsection{Drift Correction and Noise Estimation}
\label{subsec:drift}
\begin{figure}[!t]
    \centering
    \includegraphics[width=0.9\linewidth]{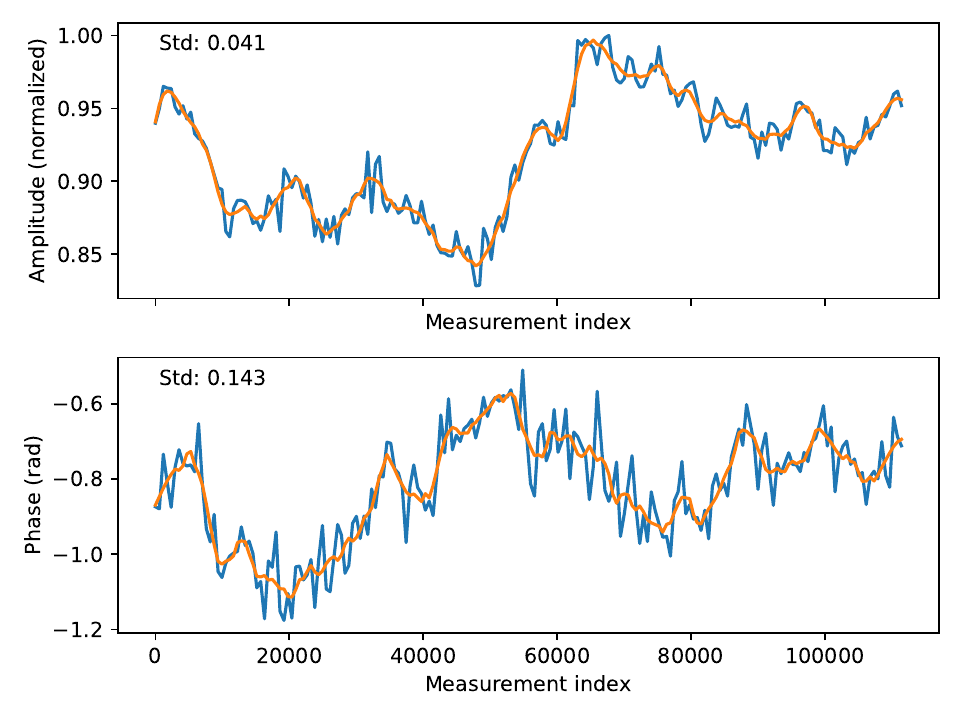}
    \caption{Drift in the DRF measurement ascertained from the periodic measurement of a reference position throughout the scan.}
    \label{fig:drift}

\end{figure}

A reference position was set in the scanning strategy to which the sources periodically returned. Repeated measurement of the reference position allowed for estimations of measurement noise. In addition, long term drifts in the source power could be corrected for by applying a scaling factor based on the drift in measured amplitude seen by the reference position measurements. The standard deviation of both amplitude and phase of the reference position measurements were taken to estimate the level of amplitude and phase noise in the setup. These values were used together with simulations of the expected DRF to estimate the noise floor above which mode sensitivities could be measured. The reference positions measurements are shown in Figure \ref{fig:drift}. These measurements showed that the 1-sigma uncertainty in the fringe amplitude was \qty{4}{\percent} and the 1-sigma phase uncertainty was \qty{0.14}{rad}.

\subsection{DRF Propagation}

Typically, the optical modes of a system are defined at the image plane. However, they may be measured in an intermediate plane and propagated to the final image plane if sufficiently sampled.In our configuration, size of the beamsplitter and scanning extent prevented us from placing the final aperture of the exterior photomixer optics (which defined the source) at an image plane. Instead, the source was position \qty{35}{\mm} behind the image plane, and the DRF was measured at this plane.

The DRF can be propagated from the plane in which the sources are
scanned to the plane in which the detector modes will be decomposed.

Propagation of a correlation matrix follows the same principles as
propagation of the individual fields, since it is a statistical property
of the fields themselves \cite{visserFourierProcessingPartially2017, mandelOpticalCoherenceQuantum, martinLongwaveOptics1993}.

The DRF was propagated from the source plane to the detector image plane using a
Fourier optics approach based on the APWS representation \cite{martinLongwaveOptics1993}.
Using this method, each source was decomposed into an angular spectrum
of plane waves, of which correlations between those plane waves were represented
by the 2D Fourier transform of the complex DRF, $\tilde{D}$ \cite{davisAnalysisTechniquesComplexfield2018},

\begin{multline}
    \tilde{D}\left(k_{1}, k_{2}\right)=\frac{1}{(2 \pi)^2} \int_{-\infty}^{\infty} \int_{-\infty}^{\infty} D (x_1, x_2)_{(z=0)} \\
    e^{-i\left(k_{1} x_1 + k_{2} x_2\right)} d x_1 d x_2,
\end{multline}

where $k_{i} = \frac{2 \pi m}{M dx_i}$ are the spatial frequencies of $D$ in Fourier space, corresponding to the plane wave propagation directions for source $i$. $M$ is the number of sampled positions of the source, and $dx_i$ is the sample spacing, and $-M/2 \le m \le M/2$.

Representing the DRF in reciprocal space conferred two major advantages, namely, a far-field filtering step could easily be applied by eliminating large angle components of the transformed DRF, and propagation of the DRF from the plane in which the sources were scanned to a focal plane of the system under test was trivial.
Propagation was achieved by multiplying each point of the transformed DRF by a phase changing
parameter along the $k_z$ propagation direction, $\exp\left(\pm i \sqrt{k^2 + k_{1}^2 - k_{2}^2} z\right)$, before applying the inverse Fourier transform to recreate the DRF at the desired propagation distance $z_0$,

\begin{multline}
    D(x_1, x_2)_{(z=z_0)} = \frac{1}{(2 \pi)^2} \int_{-\infty}^{\infty} \int_{-\infty}^{\infty} \tilde{D}\left(k_{1}, k_{2}\right) \\
    e^{ \pm i \sqrt{k^2 + k_{1}^2 - k_{2}^2} z} e^{i\left(k_{1} x_1 + k_{2} x_2\right)} d k_{1} d k_{2}.
    \label{eqn:apws_prop}
\end{multline}

Note that due to the complex conjugate in the correlation matrix
represented by the DRF, source 1 behaves as an outgoing wave, while source 2
behaves as an incoming wave \cite{mandelOpticalCoherenceQuantum}, which is represented by the difference in sign between the $k_{1}$ and $k_{2}$ terms of the propagation factor in Equation \ref{eqn:apws_prop}.

\subsection{Spatial Filtering of DRF}
\begin{figure}[!t]
    \centering
    \includegraphics[width=0.9\linewidth]{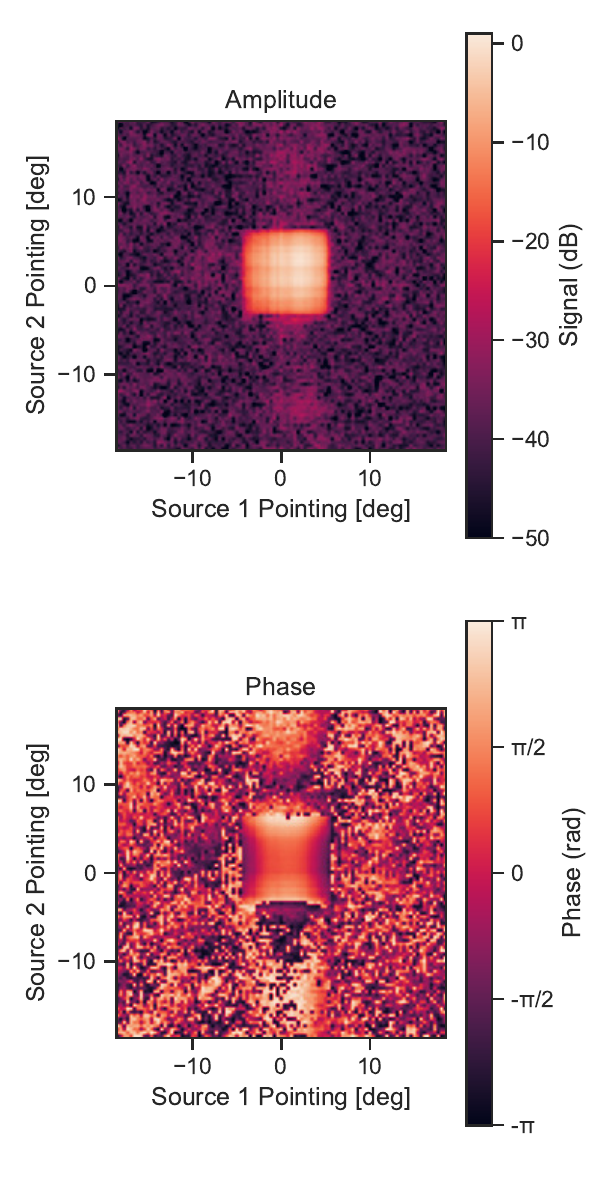}
    \caption{The Angular Plane Wave Spectrum (APWS) representation of the DRF.}
    \label{fig3}

\end{figure}

An additional filtering step in the angular spectrum domain was also
applied to limit the impact of detector noise and straylight on the DRF. 
A square filter mask was applied to the APWS shown in Figure \ref{fig3} with a side length slightly larger than the illumination angles of the detector defined by the system pupil diameter. Since the DRF is a correlation matrix of two one-dimensional beam cuts, the filter mask is square as opposed to a circularly symmetric mask which would be used in two-dimensional beam propagation methods \cite{Davis2019}.

\begin{figure}[!t]
    \centering
    \includegraphics[width=0.9\linewidth]{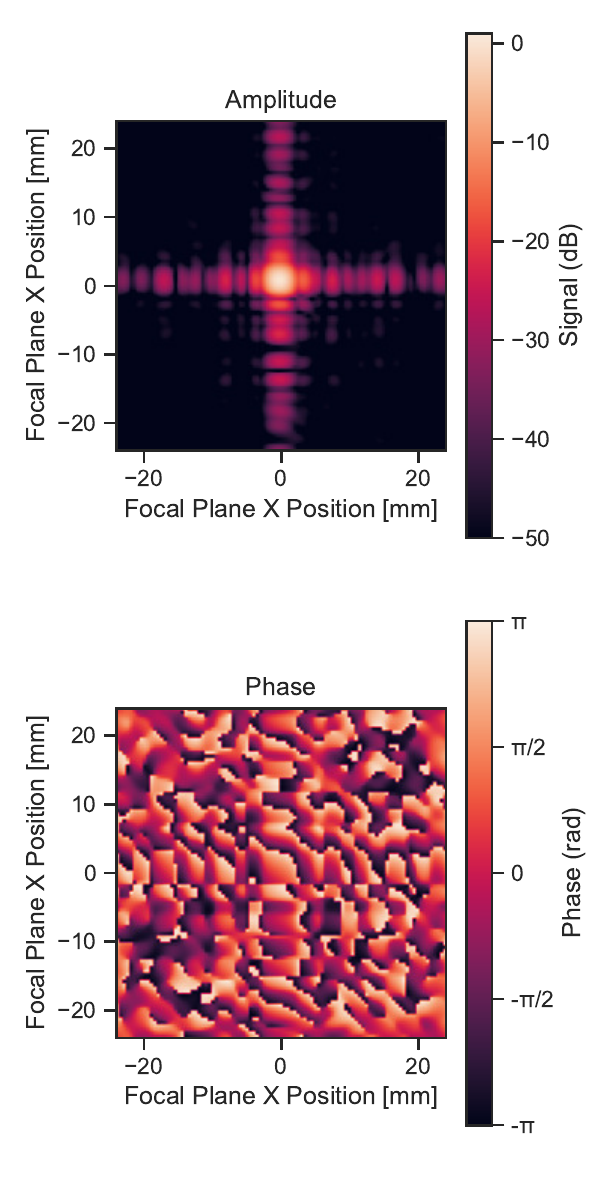}
    \caption{The DRF after spatial filtering and propagation to the focal plane.}
    \label{fig4}
\end{figure}

After propagation and spatial filtering, the DRF was reconstructed with an inverse two-dimensional Fourier transform. Figure \ref{fig4} shows the final realization of the 
DRF in the detector image plane. The final step was the diagonalization of the DRF matrix to determine the natural modes of the detector.

\subsection{DRF Decomposition}
As discussed in Section \ref{subsec:csd},
the DRF can be expressed as the cross-spectral density of a monochromatic field, which itself can be represented in the form \cite{wolfNewTheoryPartial1982}:

\begin{equation}
    D\left(\mathbf{r}_1, \mathbf{r}_2\right) = \sum_n \lambda_n \psi_n^{\ast}\left(\mathbf{r}_1\right) \psi_n\left(\mathbf{r}_2\right)
\end{equation}

On the right hand side, the functions $\psi_n(\vec{r})$ are eigenfunctions, and the coefficients $\alpha_n$ are eigenvalues, of the integral equation,

\begin{equation}
    \int D\left(\mathbf{r}_1, \mathbf{r}_2\right) \psi_n\left(\mathbf{r}_1\right) \mathrm{d}r_1 =
    \lambda_n \psi_n\left(\mathbf{r}_2\right).
\end{equation}

\begin{figure}[!t]
    \centering
    \includegraphics[width=0.9\linewidth]{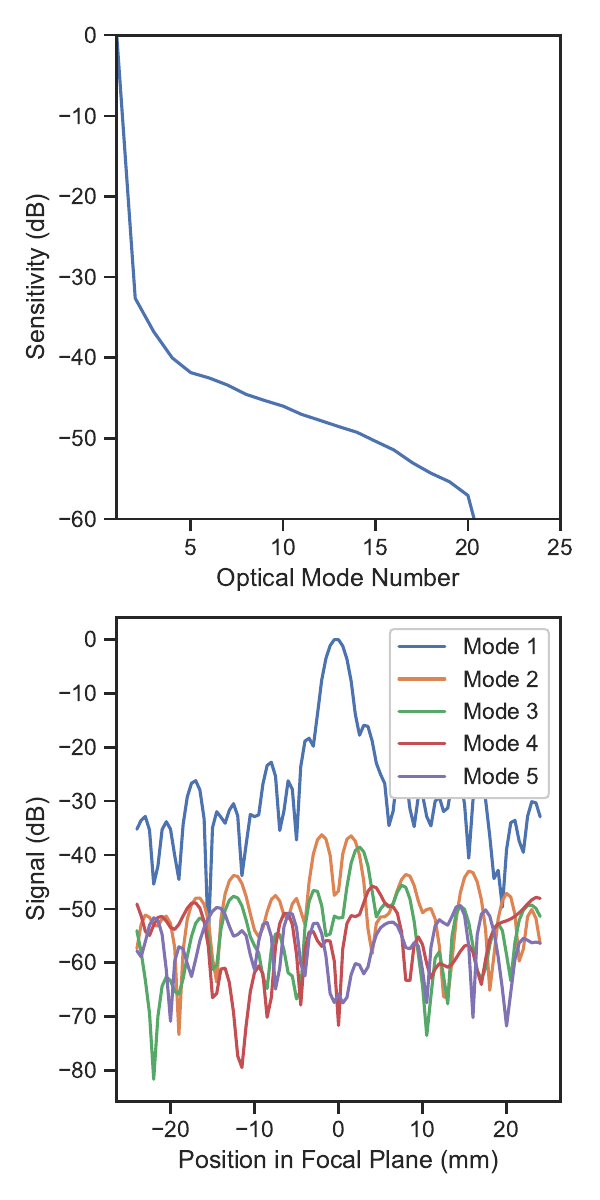}
    \caption{Eigenvalues and eigenvectors of the propagated DRF, represented as natural spatial modes and their relative sensitivities.}
    \label{fig5}

\end{figure}

The eigenfunctions are orthonormal and represent the natural optical modes of the field, and the eigenvalues are the corresponding sensitivities to each mode \cite{wolfNewTheoryPartial1982}. Therefore, diagonalization of the DRF after propagation to the desired plane directly recovers the optical modes and relative sensitivites
(Figure \ref{fig5}). In general, the matrix $D$ is not diagonalizable, therefore a pseudoinverse is computed using singular value decomposition (SVD) \cite{Withington2017}.

It was initially presumed that the second mode corresponded to the cross-polarization response of the antenna.
However, this detector is only sensitive to power dissipated in the central line of a CPW feed of the antenna. Therefore it has an intrinsic single mode response. The cross-polarization of this antenna is formed by the coherent projection and scattering of this mode. Therefore in the language of EAI, the cross-polarization mode has a high level of coherence with the co-polarization, so can only be extracted using orthogonal polarisation measurements as in \cite{Davis2019}.

Simulations of the DRF produced by the coupling of two point sources with a single-moded Gaussian beam with similar levels of added phase and amplitude noise produced a DRF with a similar decomposition.
Therefore, we concluded the higher-order modes present in this experiment are limitations of the setup based on random amplitude and phase noise of the sources (Section \ref{subsec:drift}) and represent the limit of the relative higher-order mode sensitivites that can be recovered with this setup.

The demonstration described in this work proved that this detector \cite{baselmansKilopixelImagingSystem2017} can be described as having a single dominant spatial mode and that cross-pol and surface wave stray light can be described as coherent with the main antenna response, at least within measurement accuracy. This would not be expected in a intrinsically multi-moded detector, like a distributed absorber detector such as the LEKID detector \cite{shuPrototypeHighAngular2018, adamNIKA2LargefieldofviewMillimetre2018}. Application of this technique to such detectors will be a topic of future focus, as it has implications for how they couple, particularly for application in wideband instruments such as Herschel/SPIRE and SPICA/SAFARI \cite{lapModelingPartiallyCoherent2022}.

\section{Conclusion}
Energy Absorption Interferometry is a powerful technique which reveals
not only information about the beam spatial distribution and pointing,
which would be extracted from traditional phase and amplitude
measurement techniques \cite{Davis2019}, but also the natural modes
through which the system can absorb energy. This work builds on
previously published results \cite{Moinard2017a, Thomas2012, lapModelingPartiallyCoherent2022}, which applies these techniques to different
wavelength regimes to include the use of continuously tunable
monochromatic sources, such as photomixers, and novel phase modulation
techniques using fiber stretchers. We validated that the lens-antenna
coupled devices measured are predominantly single-moded as expected,
with low level higher order modes from mode conversion in the optics.

The continuous wavelength tunability of photomixers allows the modal
composition of the optical response of the system to be measured at any
wavelength of interest. This is particularly advantageous for grating
based systems, in which a spectrum may be dispersed over an array of
similarly sized detectors where the power coupling of a detector to
optical modes propagating through the system varies significantly with
wavelength.

\section*{Acknowledgments}
The authors would like to thank David Thoen (then TU Delft) and Vignesh Murugesan of SRON for array fabrication; Jochem Baselmans (SRON/TU Delft) for array design and general support; and Ozan Yurduseven and Nuria Llombart of THz sensing group of TU Delft for the antenna design used in this work.
Stephen Yates would like to thank to Nicolas Reyes and in particular Ivan Cámara Mayorga of MPIfR Bonn for useful discussions.
This publication is part of the projects ‘Space simulator for ultra-sensitive cameras’ (614.061.609) and  ‘The SAFARI Imaging Spectrometer on the SPICA space observatory’  (184.032.209) of the research programmes NWO-M and Large Scale Rearch Infrastructure which are partly financed by the Dutch Research Council (NWO).

\bibliography{refs.bib}
\bibliographystyle{ieeetr}

\end{document}